\documentclass[12pt,reqno]{amsart}

\usepackage{amsmath,amssymb,latexsym,amscd,amsthm} 
\usepackage{graphicx}
\usepackage[square]{natbib}   

\newtheorem{theorem}{Theorem}[section]  

\theoremstyle{definition}

\theoremstyle{remark}
\newtheorem{remark}[theorem]{Remark}

\usepackage{booktabs}
\usepackage{hyperref}

\title{Computing knock out strategies in metabolic networks}
\author[Utz-Uwe Haus]{Utz-Uwe Haus}
\address{Institut f\"ur Mathematische Optimierung,
Otto-von-Guericke-Universit\"at, Universit\"atsplatz 2, D-39106 
Magdeburg, Germany}
\email{haus@imo.math.uni-magdeburg.de}
\author[Steffen Klamt]{Steffen Klamt}
\address{Max-Planck-Institut f\"ur Dynamik komplexer technischer Systeme,
Sandtorstr.~1, D-39106 Magdeburg, Germany}
\email{klamt@mpi-magdeburg.mpg.de}
\author[Tamon Stephen]{Tamon Stephen} 
\address{Department of Mathematics, Simon Fraser University,
8888 University Drive, Burnaby, British Columbia  V5A 1S6 Canada}
\email{tamon@sfu.ca}


\def\R{\mathbb{R}}     
\def\H{\mathcal{H}}     
\def\E{\mathcal{E}}     
     
\def\Tr{\operatorname{Tr}}
\def\discuss#1{}

\begin{document}

\newif\iftablestoend
\tablestoendfalse

%
%
\def\poly{\operatorname{poly}}
\def\cna{{\small\tt CellNetAnalyzer}}
\def\fa{{\small\tt FluxAnalyzer}}
\def\acet{{\tt acet}}
\def\succ{{\tt succ}}
\def\glyc{{\tt glyc}}
\def\gluc{{\tt gluc}}

%
%
\maketitle

\begin{abstract}
Given a metabolic network in terms of its metabolites and
reactions, our goal is to efficiently compute the minimal knock out 
sets of reactions required to block a given behaviour.
We describe an algorithm which improves the computation of
these knock out sets when the elementary modes (minimal functional
subsystems) of the network are given.  We also describe an algorithm
which computes both the knock out sets and the elementary modes
containing the blocked reactions directly from the description 
of the network and whose worst-case
computational complexity is better than the algorithms currently
in use for these problems.  Computational results are included.

\end{abstract}
%
%
\section{Introduction}
Systems biology studies the complex systems which occur at many levels
of biology. Such systems involve large numbers of components and
interactions. We consider {\it metabolic networks}, that is, a set of
metabolites that can be interconverted by biochemical reactions. A
fundamental question about metabolic networks is to find knock out
strategies that block the operation of a given reaction or set of reactions. 
A target reaction is 
blocked if it cannot operate in a steady state. Some reactions can
be easily knocked out, while others may be expensive or impossible to
knock out directly. In this case, we consider the problem of blocking
target reactions by inhibiting other reactions so that the targets
cannot continue in a steady state.
Some applications of this problem are outlined in~\citep{KG04}
and~\citep{Kla06}, and an implementation has been included as part of
{\cna}~\citep{Klamt}, a {\small\tt MATLAB} package for analyzing
cellular and biochemical networks.

In this paper, we consider methods of computing
the minimal sets of reactions that need to be
disabled to block a given target.  We call such a knock out  
set a {\it cut set}.  We remark that we focus only on the
(inclusion-wise) {\it minimal} cut sets since these are
the cheapest ways of blocking reactions in terms of
effort and impact on the system.  The list of minimal cut sets 
contains the same information as the full list
of cut sets but is much shorter.

We consider two main approaches.
The first is to build the hypergraph of elementary
modes and then compute the minimal cut sets as the
transversal of this hypergraph.  
This strategy has been employed successfully in \citep{KG04},
however we observe we can substantially improve the 
computation of the transversal hypergraph. 
The second approach is to compute the minimal cut sets directly
using the ideas of Gurvich and Khachiyan \citep{GK99} and others 
on generating monotone boolean formulae.  This procedure
also generates the set of elementary modes employing a given set of
reactions as a by-product, which is a potentially useful feature.

We expect that these methods can be adapted to more of the
complex systems that are typical at many levels of biology.  
Indeed, the question of finding minimal cut sets can be abstracted
to finding the minimal failure modes of a network, which
is a natural question arising in various contexts.
Some suggestions for using these methods in other
types of biochemical networks are presented in~\citep{KSRL+06}.

%
%
\section{Preliminaries}
We model a metabolic network as a number $m$ of metabolites 
involved in a set $Q$ of $q$ reactions 
(where $q$ is typically between $m$ and $2m$).  
For our purposes, these reactions can be encoded in a 
$m \times q$ matrix $N$ whose columns encode the metabolites 
produced and consumed by a given reaction.
The matrix $N$ is known as the {\it stoichiometric} matrix.
The reactions may be divided into two types: 
{\it reversible} reactions that can either
produce a given output from a given input or vice-versa;
and {\it irreversible} reactions which cannot operate in reverse.  
Let $S$ be the index set of
the reversible reactions and $U=Q \setminus S$ be the index
set of the irreversible reactions.
We call our set of target reactions $T$, for simplicity we
will usually assume they are irreversible, i.e.~$T \subseteq U$.

Given such a network, we are interested in its potential steady-state
flux vectors. In
steady state, the reaction rates balance the metabolites, i.e. for
each metabolite it holds that the sum of the rates of all reactions
consuming the metabolite equals the sum of the rates of the reactions
producing it. We can represent such a steady state as a vector $x \in
\R^q$ s.t. $Nx=0$ and $x_i \ge 0$ for all $i \in U$. Then we can
formally define a {\it cut set} $C \subset Q$ as a subset such that
the system:
\begin{equation}\label{eq:cutset}
\{x \in \R^q | Nx=0, x_i \ge 0 ~ \forall i \in U, x_c=0 ~ \forall c \in C\}
\end{equation}
has only solutions with $x_t=0$ for all $t \in T$.
A {\it minimal cut set} (MCS) is simply a cut set none of whose proper
subsets is a cut set.  

A concept closely related to minimal cuts sets is that of an
elementary mode.  An {\it elementary mode} (EM) is a minimal
set of reactions that can exist in a steady state.
The importance of EM's in metabolic networks is discussed, 
for example, in \citep{SFD00} and \citep{SKB+02}.
EM's are, up to a scaling factor, support minimal 
solutions to the system:
\begin{equation}
\{x \in \R^q | Nx=0, x_i \ge 0 ~ \forall i \in U\}
\end{equation}
The problem of computing the EM's of a given system
has a nice geometric formulation: it reduces
to finding the
extreme rays of the cone $\{r | \hat{N}r=0, r \ge 0\}$, where
$\hat{N}$ is $N$ modified to represent reversible reactions as
opposite pairs of irreversible reactions.
This is described in \citep{GK04}.  The EM's can then
be computed by applying the double description method
(see for example~\citep{FP96}) to this cone.
As observed in \citep{GK04}, EM's are
characterized up to a constant by their binary support patterns.
Hence we will also use the term EM to refer to this
support pattern, which we can view as a set of reactions.

For the purposes of finding cut sets for a given target $T$, 
we consider only the EM's
that include at least one target reaction.
Note that cut sets are exactly 
the sets of reactions that intersect each of these EM's.
The collection $\E$ of these EM's
defines a Sperner hypergraph $\H=(R,\E)$ on the set of reactions.
(A hypergraph is Sperner if it has no nested edges.)
The key observation 
is that cut sets are exactly the sets
that intersect every edge of $\H$.  
In the terminology of hypergraphs such sets are known as
{\it hitting sets} or {\it vertex covers}.  The collection of all
minimal hitting sets for $\H$ is itself a hypergraph $\H'=(R,\E')$
which is dual to $\H$ in the sense that its minimal hitting sets
are the edges of $\H'$.  The hypergraph $\H'$ is known as the 
{\it transversal hypergraph} of $\H$ and is denoted $\Tr(\H)$.

The approach to computing minimal cut sets presented in
\citep{KG04} is to first compute the EM's hypergraph $\H$
via the double description method 
and then compute $\Tr(\H)$.  The computation of $\Tr(\H)$ is
done through an enumeration scheme.
This succeeds in solving four large scale networks arising in biomass
synthesis in {\it E.coli}.  The computation benefits substantially 
from effective preprocessing, but nevertheless consumes a lot 
of time and memory.  A faster and more memory efficient algorithm
is described in Section~\ref{se:berge}.

The double description method and the algorithms of 
Section~\ref{se:berge} have uncertain complexity.  
For this reason, we give in Section~\ref{se:joint} an algorithm
which generates both the EM's and the MCS's
directly from the stoichiometric matrix which has a
surprisingly good complexity bound of $m^{\poly(\log{m})}$ in
the output size. See~\citep{stougie:07} for an overview of the known
complexity results regarding EM and MCS algorithms.


\begin{remark}
We could consider weighting the reactions and looking for
a single minimum weight cut set.  
This assumes that the costs are independent,
which is questionable - it may be possible to attain
some economies when knocking out multiple reactions.
Additionally, designing a weighting
function for blocking a metabolic reaction requires quantifying the costs to
disable various reactions, which is a labour intensive task.
Since we can in some interesting cases produce
the entire list of minimal cut sets, we view generating the
entire list as a suitable goal.  

If we do want to find a minimum weight cut set, this problem
is the {\it minimum set cover}
problem on the dual hypergraph produced by interchanging the roles 
of the vertices and edges of $\H$.  
Minimum set cover is a classical NP-complete
problem, see for example \citep{ADP80}.  
\end{remark}

\subsection{Characteristics of metabolic networks}\label{se:struct}
In our model the stoichiometric matrix $N$ is a real 
$m \times q$ matrix.  Typically, we would expect $N$ to 
have many zero entries as any particular reaction will only involve
a few metabolites.  Those non-zero entries will usually be small
integers since chemical reactions are discrete 
rearrangements of molecules.
Note that a column describes the same reaction when it
is scaled by a positive factor.  
Recall that reactions may be reversible or irreversible.
The hypergraphs of EM's and MCS's are 0-1 matrices $\H$ and $\H'$
determined by $N$.  Each row of $\H$ ($\H'$) is an indicator
function for a given EM (MCS), that is, a row indicates the
complement of a maximal $C$
such that (\ref{eq:cutset}) has a solution with $x_t>0$ for some $t \in T$
(indicates a minimal $C$ such that (\ref{eq:cutset}) has only 
solutions with $x_t=0$ for all $t \in T$).
Both $\H$ and $\H'$ have the same number of columns as $N$, 
but in our applications they will have many more rows.
The orders of the rows and columns are arbitrary, but they can
affect the performance of the algorithms.

\subsubsection{Typical behaviour} 
We understand from biological considerations that we would expect
to get many small hitting sets.  The intuition is that
such networks have some important reactions whose loss
quickly impairs the operation of the network.
This can be quantified through ``fragility coefficients'' \citep{KG04}, 
which are an average of MCS sizes.  
Their examples produced fragility coefficients in a narrow range.

\subsubsection{Test cases}\label{se:cases} 
The motivating problems from \citep{KG04} are four networks
obtained from studying biomass synthesis in {\it E.coli}.
These are the growth modes for substrates acetate, succinate,
glycerol and glucose from the network presented in \citep{SKB+02}.
The objective is to block the single target reaction representing
growth in each of these networks.
For the purposes of this computation, a pair 
of reactions corresponding to the same multifunctional
enzyme (transketolase) has been combined. 
This modifies the input hypergraph by merging a pair of
vertices, creating some nested edges.

%
%
\section{Computing minimal cut sets via elementary modes}\label{se:berge}
The method proposed in \citep{KG04} to compute
minimal cut sets involves two steps.  The first step is to compute the
set of EM's via polyhedral methods, and the second step
is to compute the transversal hypergraph of the EM's
hypergraph.  Their method of computing the transversal hypergraph
requires enumerating many possible partial 
solutions.  As a result it consumes substantial time and memory,
and is ripe for improvement.  

\subsection{Algorithms}
In this section we sketch algorithms
for the transversal hypergraph problem, including the enumeration
algorithm of \citep{KG04} and the algorithm 
described by Berge in \citep{Ber89}.
For the latter, we describe several useful modifications.

\subsubsection{Enumeration algorithm} 
This is the original algorithm implemented in {\fa} \citep{KG04},
the predecessor to {\cna}. 
Beginning at size 1, it tests for subsets of a given size.  It maintains
a list of unused partial cut sets to avoid full enumeration
of the subsets.  

The problem with this algorithm is that the list of partial cut sets
can grow quite quickly.  Nevertheless, it can solve large problems.
A major reason for this is the
abundance of small cut sets, which keeps the list of partial cut sets
manageable.

\subsubsection{Berge's algorithm}\label{se:ourberge} 
This algorithm~\citep{Ber89} orders the edges $e_1,e_2,\ldots,e_r$
of the hypergraph $\H$, and then computes in sequence
the transversal of each hypergraph $\H_i$ consisting
of edges $e_1$ through $e_i$.  This can be done by taking all
the edges created by adding a vertex from $e_i$ to an edge in
$\H_{i-1}$ and keeping all the inclusionwise-minimal edges.
$\H_1$ has an edge consisting of a single vertex for
each vertex from $e_1$.

The performance of this algorithm depends on the size of the
intermediate transversals generated, which in turn depends on the
order of the vertices. 
Intuitively we do not expect the size of a transversal
of a subgraph to substantially exceed the larger of the size of the
initial graph and the size of the final transversal hypergraph.
In practice this does often turn out to be the case, but Takata
exhibited an example where an intermediate transversal will have
size $\Theta(m^{\log(\log(m))})$ in the combined size of the input and
output for any ordering of the edges, see \citep{Hage07}.  We
do not know of any non-trivial upper bounds for the size of
intermediate transversals or how to produce favourable edge
orderings when they exist.

A naive implementation of Berge's algorithm will be slow,
but there are a number of modifications that can make it more 
effective.  
The main bottleneck is the removal of superset rows from the
list generated from $\H_{i-1}$ and $e_i$.  We do this in time
$O(n^2)$ in the length of the list using the simple algorithm
described below.
There are algorithms for superset removal that work in time
$O(n^2/\log{n})$ or slightly better, see \citep{Pri95} and \citep{SE96}.
It is known, that there is a lower bound of
$O(n^2/\log^2 n)$, if the complete subset lattice
is constructed by algorithm, see~\citep{Pri99}.
It appears that no non-trivial lower bounds are 
known for superset removal alone, and we remark that it is an interesting 
question.

When implementing Berge's algorithm, we can avoid generating
some edges that are clearly supersets before entering the removal
phase.  For instance whenever an edge $f$ of $\H_{i-1}$ intersects $e_i$,
we can add only that edge to the list of candidates for $\H_i$
since $f$ will be generated as an edge using the common vertex with
$e_i$ and all edges generated using the other vertices of $e_i$ will
contain this edge.  Additionally, we can see that the list of 
edges retained in this way will itself be superset free
since it is a subhypergraph of $\H_{i-1}$.
Hence we only need to check which of the new edges generated
are supersets of these retained edges: newly generated edges cannot
be subsets of edges from $\H_{i-1}$ since they are edges from 
$\H_{i-1}$ with an additional vertex, and if 
$f_1 \cup v_1 \subset f_2 \cup v_2$ in the new edge list, then
$f_1$ must contain $v_2$ and hence would in fact be in the retained
edges list.

Our implementation of Berge's algorithm is now included in {\cna}. 

\subsubsection{Others}\label{se:others} 
Several algorithms have recently been proposed which build on
the Berge algorithm and are effective for some problems,
see for example \citep{BMR03} and \citep{KS05}.  
A nice recent survey of methods for computing transversals is \citep{EMG06}.
It focuses on work based on the ideas of
Fredman and Khachiyan \citep{FK96} concerning generation of 
logical formulae.  These methods provide better theoretical
performance and we consider them in Section~\ref{se:joint}.

\subsection{Implementation} \label{se:impb} 
While Berge's algorithm is well known, naive implementations
of it are quite slow.  Modifications of Berge's algorithm,
such as the ones mentioned in Section~\ref{se:others} have
apparently yielded good results, but we know of no public
implementations of such an algorithm.  
Thus we implemented our own version of the algorithm described in
Section~\ref{se:ourberge}.  We used {\small\tt MATLAB} because it is the 
platform for {\fa} and {\cna}. 
Our code is freely available for academic use \citep{BergeCode}.

We tested our code on the examples of \citep{KG04}.
These are obtained from the growth-related EM's calculated in
\citep{SKB+02} via simple modifications described
in Section~\ref{se:pre}.
Our results are presented in Section~\ref{se:compb}.
Below we describe some details of our implementation, as
well as the original {\fa} implementation.
We used as much of the {\fa} setup as possible,
including input data structures and pre- and post-processing code
to facilitate comparison between the core algorithms.

\subsubsection{Preprocessing}\label{se:pre}
Beginning from the computed full set of EM's, we first select only
those containing at least one of the target reactions.
As mentioned in Section~\ref{se:struct}, some groups of
reactions may be catalyzed by the same multifunctional enzyme.
These reactions are cut simultaneously by disabling such an enzyme. 
We combine the reactions corresponding to such groups by a logical
``or'' operation as the corresponding elementary mode is 
disabled if any of its constituent reactions are disabled.
This merges vertices in the input hypergraph;
the merged vertices in the new modes represent the set of
reactions blocked by the enzyme.
In our examples there is only one target reaction (biomass synthesis)
and one multifunctional enzyme (transketolase). 
This gives us our initial hypergraph $\H$ of modes that include the target
reaction.

There are several further 
preprocessing steps applied to $\H$. 
The first is to scan for zero rows and columns, which should not occur.
The second is to find columns of ones,
which correspond to cut sets of size one and can be
noted and removed until postprocessing.  
Next duplicate columns are identified.
They are treated as a single column 
and then reexpanded during post-processing.
In terms of the hypergraph, they represent vertices that
are in exactly the same set of edges, and can thus be merged
during the calculation.  

Finally, for Berge's algorithm, 
we remove rows that are supersets of other rows.  
Our current implementation does this in time $\Omega(m^2)$. 
Superset or duplicate rows may be introduced when merging vertices
corresponding to multifunctional enzymes.
In our examples, about 20\% of the rows in $\H$ are supersets 
of other rows, these come from the pairs of EM's that contain
only one of the transketolase reactions, each pair contains one
EM using transketolase1 and transaldo, and one EM instead using
transketolase2.  The first of these becomes a superset of the
second upon merging the two columns.
Removing these supersets in preprocessing cut the observed 
running time by about 30\%.
Removing superset rows does not speed up the algorithm in
{\fa}, whose bottleneck is generating the possible cut sets
of a given size.

\subsubsection{Postprocessing}\label{se:post}
Following the application of either algorithm to the 
hypergraph produced in Section~\ref{se:pre}, there is
a small amount of postprocessing that needs to be done.
Merged vertices are reexpanded: each cut set containing
a merged vertex $v$ will be replaced by cut sets containing
exactly one of the vertices that were merged into $v$.
Size one cut sets will be introduced for each column of
ones that was removed prior to the calculation.  
Finally, the cut sets involving reactions corresponding
to a multifunctional enzyme are split into cut sets
containing the constituent reactions.

\subsubsection{Coding issues} 
Because we are working with large matrices, memory use
is a key consideration, it is essential to use a bit-level
representation of the binary matrices describing 
the EM's and MCS's.  This is done in the implementation 
of \citep{KG04}.  

We had to accommodate {\small\tt MATLAB}'s strengths and
weakness: we obtained substantial time savings through
small changes in the main bottleneck routine.
This routine removes the rows from one list that 
are supersets of rows from another.  
Since memory allocation in {\small\tt MATLAB} is slow rather than 
resizing the matrix when superset rows are identified, 
we mark rows for removal in a pass through the matrix 
and then we generate a single new matrix containing 
those rows.  It is also useful to take advantage of
{\small\tt MATLAB}'s internal parallelization.  {\small\tt MATLAB} can quickly
check a single row for super- or sub-setness against an
entire matrix using a couple of bitwise comparisons.
We always cycle through the rows of the shorter of the 
two lists and check its rows against the longer list.

\subsubsection{Computational results}\label{se:compb} 
In this section we compare the performance of the transversal
hypergraph code written for {\fa} to our implementation
of Berge's algorithm now included in {\cna}.  
Our test base is the four EM 
problems from~\citep{KG04} (see Section~\ref{se:cases})
and their hypergraph transversals.
The transversals are denoted by primes.

We first compare the algorithms from the point of view of the
largest intermediate lists generated.  This tells
us how much memory each algorithm uses, and gives an idea of
how fast it can run.  In the case of the {\fa} algorithm,
the measure is the largest list of partial cut sets generated.
In the case of the Berge algorithm, the measure is the largest
intermediate transversal generated before removing nested subsets.
Due to the reductions of Section~\ref{se:pre}, this sometimes turned
out to be smaller than output it generated after postprocessing.
Results are in Table~\ref{ta:bsizes}.  We used as inputs both the
EM's found for the networks described in Section~\ref{se:cases}
and the dual hypergraphs containing the MCS's that we computed
(denoted with a ').

\iftablestoend
\else
\begin{table}[ht!]
\begin{tabular}{l@{\quad}rr@{\quad}rr@{\quad}rr@{\quad}rr}
  \toprule
  Problem:          &  \acet &  \acet'&   \succ &       \succ'&   \glyc  &      \glyc' &    \gluc & \gluc' \\
  \midrule
  Input columns  &    104 &    104 &     104 &         104 &      105 &         105 & 105 & 105 \\
  Preprocessed columns &  21   &  98     &  26     &   101       &  28      &   103        &   34      & 103 \\
  Input rows        &    363 &    245 &  3\,421 &      1\,255 &   9\,479 &      2\,970 &  21\,592 & 4\,225 \\
  Preprocessed rows &    289  &  244 &  2\,722 &     1\,254 &   7\,472 &      2\,969 &  18\,481 & 4\,224 \\
  Raw output rows   &     54 &    280 &     159 &      2\,589 &      376 &      7\,047 & 918 & 18\,481 \\
  Final output rows       &    245 &    289 &  1\,255 &      2\,722 &   2\,970 &      7\,472 &   4\,225 & 18\,481 \\
  {\tt FluxAnalyzer} largest  & 3\,563 &     -- & 69\,628 &          -- & 342\,025 &          -- & 902\,769 & -- \\
  Berge largest     &     94 &    296 &     304 &      2\,669 &      657 &      7\,047 &   1\,714 & 18\,569 \\
    {\tt {\fa}} time & 5.1   & --    & 633.5 & -- & 8\,696.2 & -- & 54\,099.1 & -- \\
    Berge time & 0.7   & 1.1    &   7.1 &  35.8 &   29.6 & 215.0    & 206.5   & 727.4 \\
  \bottomrule
\end{tabular}
\vspace{.2\baselineskip}
\caption{Sizes of intermediate lists generated in computing transversals
and computation times.}
  \label{ta:bsizes}
  \label{ta:btimes}
\end{table}
\fi

We record the number of columns and rows both before and after
preprocessing.  The preprocessing reduces the problem size substantially
mainly by removing columns corresponding to reactions which form cut 
sets by themselves.
The number of output rows is given before and after postprocessing. 

The {\fa} routine was not able to solve the dual problems due 
to memory limitations.  This routine is much better at converting
EM graphs to MCS graphs than vice versa because many of the MCS's are
small in these instances.
For the given examples, all the EM's are large, so the computation
begins by building a long list of partial cut sets.

While Berge's algorithm provides no complexity guarantees, it worked
very well for these problems.  Of particular note is that, unlike
the {\fa} algorithm, the information carried by Berge's algorithm
during its intermediate stages was never much larger than the
size of the final output prior to preprocessing.

We also give running times in seconds in Table~\ref{ta:btimes}.  
Both codes are written in {\small\tt MATLAB}, take identical inputs, and 
use the same preprocessing code as noted in Section~\ref{se:pre}. 
These comparisons were run on a Sun Fire V890 with 32 GB
memory and 16 1200 MHz processors.  
Note that running times include some preprocessing, such as
removing duplicate rows for Berge, but excludes postprocessing.  
This method of reporting was used in \citep{KG04}.

%
%
\section{Computing Minimal Cut Sets Directly}\label{se:joint}
In this section, we consider methods of generating the
MCS's directly from the stoichiometric matrix.
The techniques outlined here are based on an algorithm of 
Fredman and Khachiyan \citep{FK96} for dualizing boolean
functions.  They offer better complexity guarantees than
the algorithms of Section~\ref{se:berge}.  They work
directly from the stoichiometric matrix for a network
and generate the hypergraph of EM's containing the blocked
reactions as a byproduct of the computation.

\subsection{Algorithm}
The cut sets generated by a given stoichiometric
matrix define a boolean function, that is a function
that takes a binary pattern of included reactions as input, and
yields 1 if this set of reactions is a cut set, and 0 
if it is not.  Further, this is a {\it monotone} function in the sense
that if a given set is a cut set, then any superset of that set is
also a cut set.  Thus the problem of finding {\it minimal} cut sets can
be viewed as a problem of representing such a boolean function via
its minimal true assignments.  The support of an EM 
is then the complement of a maximal false assignment.  

A monotone boolean function can be represented uniquely
both by its minimal true
assignments and its maximal false assignments.  The process of
converting from one representation to the other is sometimes 
called dualization, and is equivalent to the hypergraph
transversal problem.
Fredman and Khachiyan \citep{FK96} proposed an algorithm that
generates the transversal incrementally using an algorithm
that is slightly superpolynomial in the size of the graph
and the transversal.  This key idea in this algorithm is to
recurse on a variable that occurs with relatively high frequency.

As described in \citep{GK99}, this algorithm can be implemented from
a function evaluation oracle, producing both the hypergraph of
minimal true assignments and its transversal in $m^{o(\log(m))}$ 
oracle calls, where $m$ is the combined size of the
two hypergraphs.  
This fits very well with our problem: 
given the stoichiometric matrix, 
we want to generate both the MCS's 
and the EM's. 
Note that the boolean function characterizing the cut set is monotone
because every superset of a cut set is again a cut set.

We remark that the amount of memory required to generate
the next clause is bounded by $m^{\poly(\log{m})}$,
while the worst-case memory blow up for Berge is unknown,
but not polynomial.

\subsection{Implementation}\label{se:impj} 
The algorithm of \citep{FK96} gives remarkable theoretical results,
but the only implementation we know of is that of \citep{BEGK06}.
Their code is not public and uses hard-coded oracles different from
the one for our problem.
So we implemented a prototype of the Fredman-Khachiyan 
algorithm with a suitable oracle, again in {\small\tt MATLAB} for easy 
comparison to {\fa} and the results in Section~\ref{se:impb}.
We emphasize, though, that this algorithm proceeds directly
from the stoichiometric matrix, in contrast to the Berge
algorithm, which requires the completed computation of the EM's
as input.

\subsubsection{Oracle}\label{se:oracle} 
We test whether a given set $C$ is a cut set by checking 
if the system (\ref{eq:cutset}) has any solutions with
$x_t > 0$ for some $t \in T$.
This is a linear programming feasibility problem, and thus
can be solved in polynomial time.  
We do this via an external call to
{\small\tt CPLEX}~\citep{ilog-cplex-uuh}, which is known to have a fast and reliable LP solver.
We can test for non-trivial solutions by
maximizing the sum of the target variables 
$\sum_{i \in T} x_i$ subject to (\ref{eq:cutset}).
If this is greater than zero or unbounded we have a non-trivial
solution.

\subsubsection{Duality checker}\label{se:checker} 
Using this oracle, we implemented a version of the ``Algorithm A''
duality checker from \citep{FK96}.  This checker either verifies 
that our current collections of EM's and MCS's form dual hypergraphs,
in which case both sets are complete and we are done, or it
finds a set of reactions that is not a superset of any 
current EM, and whose complement is not a superset of any 
current MCS.  Given such a clause we use the oracle to check if it 
is a mode or if its complement is a cut set.  If it is a mode, we
test whether it remains a mode upon removing in turn each of its 
constituent reactions 
-- if the resulting set is no longer a mode, we return
the removed reaction to the set, otherwise it stays out.  
In this way we obtain an EM, which we add to our list.
Similarly, given a cut set, we obtain an MCS.

The essence of the algorithm is to recurse on a frequently
occurring reaction in one of the lists until we arrive at
a trivial case.  The recursion then tests separately if
duality holds assuming that this variable is true and assuming
that it is false.  By taking a frequently occurring
variable, Fredman and Khachiyan ensure that the sizes of
the lists decrease fast enough to guarantee that the
algorithm runs in time $m^{O(\log^2(m))}$, where $m$ is
the current joint length of the two lists.
As with Berge's algorithm, we find that the bottleneck is
removing supersets from lists: when we recurse on a reaction
some of the elements of the reduced lists, which omit this
reaction, will no longer be minimal.  These must be removed,
and this takes $O(m^2)$ time.  

Fredman and Khachiyan also provide an ``Algorithm B'' which 
achieves further economy through observing some interdependencies 
in the two subproblems.  This reduces the time guarantee to
$m^{o(\log(m))}$, but the resulting algorithm is much
more complicated, so we did not implement a prototype.

We did try several variations of the simpler Algorithm A.
We found some useful corners to cut: it is helpful to
short circuit the recursion by treating more base cases
than suggested by the algorithm.  We can substantially
reduce the number of superset removal calls required
by doing them only when necessary before recursing rather
than at the start of the checking routine.

We ran our code on the stoichiometric matrices that
are used to generate EM's in \citep{SKB+02}.
Our oracle tests if a set of reactions blocks the growth reaction,
the two transketolase reactions are treated as a single reaction for
the purposes of blocking.   Thus the EM's we generate use 
a single bit to indicate if either of the two reactions are active.
As with the preprocessing of Section~\ref{se:pre} this merges
certain EM's.

\iftablestoend
\else
\begin{table}[ht!]
  \begin{tabular}{l@{\quad}rrrr}
    \toprule
    Problem     &\multicolumn{1}{c}{\acet}& \multicolumn{1}{c}{\succ} & \multicolumn{1}{c}{\glyc} & \multicolumn{1}{c}{\gluc} \\
    \midrule
    EM's        & 289 & 2\,722 &  7\,472 & 18\,481 \\
    MCS's       & 245 & 1\,255 &  2\,970 & 4\,225 \\
    \addlinespace
    Total recursive calls & 107\,781 & 11\,129\,110 & 122\,136\,668 & 764\,239\,195 \\
    Time to generate & 194.8 & 10\,672.2 & 103\,511.2 & 677\,599.3 \\
    \hline
  \end{tabular}
  \vspace{.2\baselineskip}
  \caption{Call counts for the Fredman-Khachiyan algorithms.}
\label{ta:jsizes}
\end{table}
\fi

In Table~\ref{ta:jsizes}
we record the number of calls
to the (recursive) duality checker used in our implementation
of Fredman and Khachiyan's algorithm. 
Each stoichiometric matrix has 89 metabolites and 105 or 106
reactions.  

The algorithm is written in {\small\tt MATLAB} and the oracle uses external
calls to {\small\tt CPLEX} when necessary to solve linear programs.
Running times are included in Table~\ref{ta:jsizes} above.
We used the same computer as in Section~\ref{se:compb}.

We remark that if our objective is to produce the
EM's containing the target reactions rather than the MCS's, 
we can compress the network as described in \citep{KGv06}.  
This speeds up the computation, since it now needs to generate
only the few cut sets for the compressed modes 
(which are hard to interpret), see Table~\ref{ta:compress_gen}.
In this case, since our objective is to produce the EM's,
we did not merge the two transketolase reactions.

\iftablestoend
\else
\begin{table}[ht!]
 \begin{tabular}{l@{\quad}rrrrrc}
   \toprule
   Problem & \acet & \succ & \glyc & \gluc \\
   \midrule
    Compressed reactions   & 40  & 40     &       42 & 42 \\
    EM's        & 363 & 3\,421 &  9\,479 & 21\,592 \\
    Total recursive calls & 38\,503 & 3\,487\,200 & 20\,971\,005 & 217\,252\,316 \\
    Time to generate & 45.8 & 2\,707.0 & 17\,202.0 & 210\,749.8 \\
   \bottomrule
 \end{tabular}
 \vspace{.2\baselineskip}
 \caption{Data for generating the EM's from the compressed network}
\label{ta:compress_gen}
\end{table}
\fi

In fact it is possible to do most of this compression in such a way
that the cut sets generated can be expanded to yield the MCS's for
the original system.  Working with these partially compressed networks
takes less than 10\% longer than the fully compressed networks reported
in Table~\ref{ta:compress_gen}.

We can also produce the full set of EM's using this
method by blocking the full set of reactions.  
This requires modifying the oracle to check for non-trivial
solutions to (\ref{eq:cutset}) via a rank check.
If we generate them from the compressed matrices, this is
somewhat tractable, but slower than generating only those
containing the target reactions.

\section{Conclusions}
As has been often seen in recent years, biological problems are
ripe for the application of mathematics.  
In Section~\ref{se:berge}, we use a non-trivial, but simple
algorithm to compute MCS's from EM's
in minutes rather than hours in the context of a large metabolic
network problem.  With such large data sets a careful implementation
of the algorithm was as essential to make it useful.

While Berge's algorithm is successful in practice, it
is poorly understood in theory, and thus must be considered
suspect.  Additionally, it requires the precomputation of the EM's 
via an algorithm which also has uncertain worst-case complexity.
Thus we also considered algorithms based on the
dual generation framework of Fredman and Khachiyan \citep{FK96}.
These offer a guaranteed theoretical performance which is
close to polynomial, i.e. $m^{\poly(\log{m})}$ where $m$ is
the joint size of the EM's and MCS's.  

The Fredman-Khachiyan oracle-based algorithms also have the 
advantage that, unlike Berge's algorithm, the lists are
generated {\it incrementally} - at each step a new EM or MCS is
added to the current collection, and in time and
memory $m^{\poly(\log{m})}$
in the current joint size of the lists.  
Thus if we only have
the resources to do a partial computation, we will get a 
partial answer.  Indeed, much larger systems exists
for which the full sets of EM's and MCS's are too large
to store.  In this situation, even if we could compute a
partial list of EM's, its dual is meaningless.
In contrast, the oracle-based algorithm can produce a
sampling of EM's and MCS's as resources permit.

The oracle-based algorithms also provide a method of computing
the EM's containing target reactions without computing the
full set of EM's.  This is desirable since there may be an
enormous number of EM's, only a small fraction of which
contain the target reactions.  The double description algorithm
can be modified to compute only the EM's containing a target,
however as noted in \citep{KGv06}, unless implemented carefully,
this may be slower than computing the full set of reactions.

However, there are several clear drawbacks to oracle-based
algorithms.  The obvious ones are that they are more difficult
to implement and slower.  The high number of recursive calls 
used to solve the small {\acet} problem 
(see Table~\ref{ta:jsizes}) suggests that it will be difficult
to make such an algorithm competitive with Berge, especially
in the case where the EM's are given.  It is encouraging
that our simple implementation is able to solve even the
largest problem ({\gluc}), although the time required was
long.  This gives us some hope that a more advanced implementation 
of the oracle-based algorithm could be competitive in this
application.  One place to start would be to implement
Fredman and Khachiyan's more intricate, but theoretically
faster $m^{o(\log(m))}$ algorithm.

The oracle-based methods could be improved
by additional preprocessing as is done when obtaining the EM's 
from the MCS's.  For example, single reaction cut sets are
treated separately in the algorithms of Section~\ref{se:berge},
this could also be implemented in the Fredman-Khachiyan 
framework.  
We expect this  would yield a mild improvement in the running time.

\subsection{Acknowledgments}
We are grateful for fruitful discussions with Annegret Wagler and
Robert Weismantel
This work was partially supported by
DFG FG-468, 
the Research Focus Program Dynamic Systems funded by the
Kultusministerium of Saxony-Anhalt, and by 
a President's Research Grant at Simon Fraser University.

%
%
\bibliographystyle{amsalpha}

\begin{thebibliography}{CLMS{\etalchar{+}}07}

\bibitem[ADP80]{ADP80}
G.~Ausiello, A.~D'Atri, and M.~Protasi, \emph{{Structure preserving reductions
  among convex optimization problems}}, J. Comput. Syst. Sci. \textbf{21}
  (1980), 136--153.

\bibitem[BEGK06]{BEGK06}
Endre Boros, Khaled Elbassioni, Vladimir Gurvich, and Leonid Khachiyan,
  \emph{{An Efficient Implementation of a Quasi-Polynomial Algorithm for
  Generating Hypergraph Transversals and its application in Joint Generation}},
  Discrete Applied Math \textbf{154} (2006), no.~16, 2350--2372.

\bibitem[Ber89]{Ber89}
Claude Berge, \emph{{Hypergraphs. Combinatorics of finite sets. Transl. from
  the French}}, {North-Holland Mathematical Library, 43. Amsterdam etc.:
  North-Holland. x, 255 p.}, 1989.

\bibitem[BMR03]{BMR03}
James Bailey, Thomas Manoukian, and Kotagiri Ramamohanarao, \emph{{A Fast
  Algorithm for Computing Hypergraph Transversals and its Application in Mining
  Emerging Patterns}}, {Proceedings of the 3rd IEEE International Conference on
  Data Mining (ICDM)}, {{IEEE} Computer Society}, 2003, pp.~485--488.

\bibitem[CLMS{\etalchar{+}}07]{stougie:07}
Flavio Chierichetti, Vincent Lacroix, Alberto Marchetti-Spaccamela,
  Marie-France Sagot, and Leen Stougie, \emph{Modes and cuts in metabolic
  networks: {C}omplexity and algorithms}, Tech. Report Ext. rep. 2007-01,
  Technische Universiteit Eindhoven, Eindhoven, 2007.

\bibitem[EMG07]{EMG06}
Thomas Eiter, Kazuhisa Makino, and Georg Gottlob, \emph{{Computational aspects
  of monotone dualization: a brief survey}}, Discrete Applied Math,
  doi:10.1016/j.dam.2006.04.017, in press., 2007.

\bibitem[FK96]{FK96}
Michael~L. Fredman and Leonid Khachiyan, \emph{On the complexity of dualization
  of monotone disjunctive normal forms}, J. Algorithms \textbf{21} (1996),
  no.~3, 618--628.

\bibitem[FP96]{FP96}
Komei Fukuda and Alain Prodon, \emph{{Double Description Method Revisited}},
  {Deza, M., Euler, R. and Manoussakis, I. (eds.) Combinatorics and Computer
  Science, Springer. Lect. Notes Comput. Sci. 1120}, Springer, 1996,
  pp.~91--111.

\bibitem[GK99]{GK99}
V.~Gurvich and L.~Khachiyan, \emph{{On generating the irredundant conjunctive
  and disjunctive normal forms of monotone Boolean functions.}}, Discrete
  Applied Math. \textbf{96-97} (1999), 363--373.

\bibitem[GK04]{GK04}
Julien Gagneur and Steffen Klamt, \emph{{Computation of elementary modes: a
  unifying framework and the new binary appraoch}}, BMC Bioinformatics
  \textbf{5} (2004), no.~175.

\bibitem[Hag07]{Hage07}
Matthias Hagen, \emph{{Lower bounds for three algorithms for the transversal
  hypergraph problem}}, 33rd International Workshop on Graph-Theoretic Concepts
  in Computer Science (WG 2007), 2007, to appear.

\bibitem[HKS07]{BergeCode}
Utz-Uwe Haus, Steffen Klamt, and Tamon Stephen, \emph{{MATLAB code for
  hypergraph transversal}}, Available at: {\tt \small
  http://www.math.sfu.ca/{\textasciitilde}tamon/Software/Berge/index.html},
  2007.

\bibitem[ILO07]{ilog-cplex-uuh}
ILOG, \emph{{\small CPLEX}}, 1997--2007, For information see: {\small \tt
  {http://www.ilog.com/products/cplex/}}.

\bibitem[KG04]{KG04}
Steffen Klamt and Ernst~Dieter Gilles, \emph{{Minimal cut sets in biochemical
  reaction networks}}, Bioinformatics \textbf{20} (2004), no.~2, 226--234.

\bibitem[KGvK06]{KGv06}
Steffen Klamt, Julien Gagneur, and Axel von Kamp, \emph{Algorithmic approaches
  for computing elementary modes in large biochemical reaction networks}, IEE
  Proceedings Systems Biology \textbf{154} (2006), no.~4, 249--255.

\bibitem[Kla06]{Kla06}
Steffen Klamt, \emph{{Generalized concept of minimal cut sets in biochemical
  networks}}, Biosystems \textbf{83} (2006), no.~2-3, 233--247.

\bibitem[KS05]{KS05}
Dimitris Kavvadias and Elias~C. Stavropoulos, \emph{{An Efficient Algorithm for
  the Transversal Hypergraph Generation}}, Journal of Graph Algorithms and
  Applications \textbf{9} (2005), no.~2, 239--264.

\bibitem[KSRG07]{Klamt}
Steffen Klamt, Julio Saez-Rodriguez, and Ernst~Dieter Gilles, \emph{{Structural
  and functional analysis of cellular networks with {\tt CellNetAnalyzer}}},
  BMC Systems Biology \textbf{1:2} (2007), CellNetAnalyzer can be downloaded
  for free (academic use) via the following web-site: \\ {\tt \small
  http://www.mpi-magdeburg.mpg.de/projects/cna/cna.html}.

\bibitem[KSRL{\etalchar{+}}06]{KSRL+06}
Steffen Klamt, Julio Saez-Rodriguez, Jonathan Lindquist, Luca Simeoni, and
  Ernst~Dieter Gilles, \emph{A methodology for the structural and functional
  analysis of signaling and regulatory networks}, BMC Bioinformatics \textbf{7}
  (2006), no.~56.

\bibitem[Pri95]{Pri95}
Paul Pritchard, \emph{A simple sub-quadratic algorithm for computing the subset
  partial order}, Inform. Process. Lett. \textbf{56} (1995), no.~6, 337--341.

\bibitem[Pri99]{Pri99}
\bysame, \emph{On computing the subset graph of a collection of sets}, J.
  Algorithms \textbf{33} (1999), no.~2, 187--203.

\bibitem[SE96]{SE96}
Hong Shen and D.~J. Evans, \emph{Fast sequential and parallel algorithms for
  finding extremal sets}, Int. J. Comput. Math. \textbf{61} (1996), no.~3-4,
  195--211.

\bibitem[SFD00]{SFD00}
Stefen Schuster, David~A. Fell, and Thomas Dandekar, \emph{{A general
  definition of metabolic pathways useful for systematic organization and
  analysis of complex metabolic network}}, {Nat.~Biotechnol.} \textbf{18}
  (2000), 326--332.

\bibitem[SKB{\etalchar{+}}02]{SKB+02}
J{\"o}rg Stelling, Steffen Klamt, Katja Bettenbrock, Stefan Schuster, and
  Ernst~Dieter Gilles, \emph{Metabolic network structure determines key aspects
  of functionality and regulations}, Nature \textbf{420} (2002), 190--193.

\end{thebibliography}

\newcommand{\etalchar}[1]{$^{#1}$}
\providecommand{\bysame}{\leavevmode\hbox to3em{\hrulefill}\thinspace}
\providecommand{\MR}{\relax\ifhmode\unskip\space\fi MR }
\providecommand{\MRhref}[2]{%
  \href{http://www.ams.org/mathscinet-getitem?mr=#1}{#2}
}
\providecommand{\href}[2]{#2}

\end{document}